\documentclass[12pt]{article}
\usepackage{epsf}
\usepackage{amsmath}
\def\hybrid{\topmargin 0pt      \oddsidemargin 0pt
	\headheight 0pt \headsep 0pt
	\textheight 9in         
	\textwidth 6.25in       
	\marginparwidth .875in
	\parskip 5pt plus 1pt   \jot = 1.5ex}

\catcode`\@=11
\def\marginnote#1{}
\newcount\hour
\newcount\minute
\newtoks\amorpm
\hour=\time\divide\hour by60
\minute=\time{\multiply\hour by60 \global\advance\minute by-\hour}
\edef\standardtime{{\ifnum\hour<12 \global\amorpm={am}%
	\else\global\amorpm={pm}\advance\hour by-12 \fi
	\ifnum\hour=0 \hour=12 \fi
	\number\hour:\ifnum\minute<10 0\fi\number\minute\the\amorpm}}
\edef\militarytime{\number\hour:\ifnum\minute<10 0\fi\number\minute}

\def\draftlabel#1{{\@bsphack\if@filesw {\let\thepage\relax
   \xdef\@gtempa{\write\@auxout{\string
      \newlabel{#1}{{\@currentlabel}{\thepage}}}}}\@gtempa
   \if@nobreak \ifvmode\nobreak\fi\fi\fi\@esphack}
	\gdef\@eqnlabel{#1}}
\def\@eqnlabel{}
\def\@vacuum{}
\def\draftmarginnote#1{\marginpar{\raggedright\scriptsize\tt#1}}

\def\draft{\oddsidemargin -.5truein
	\def\@oddfoot{\sl preliminary draft \hfil
	\rm\thepage\hfil\sl\today\quad\militarytime}
	\let\@evenfoot\@oddfoot \overfullrule 3pt
	\let\label=\draftlabel
\let\marginnote=\draftmarginnote
	\let\marginnote=\draftmarginnote
   \def\@eqnnum{(\theequation)\rlap{\kern\marginparsep\tt\@eqnlabel}%
\global\let\@eqnlabel\@vacuum}  }

\def\numberbysection{\@addtoreset{equation}{section}
\def\theequation{\thesection.\arabic{equation}}}

\def\underline#1{\relax\ifmmode\@@underline#1\else
	$\@@underline{\hbox{#1}}$\relax\fi}

\def\titlepage{\@restonecolfalse\if@twocolumn\@restonecoltrue\onecolumn
     \else \newpage \fi \thispagestyle{empty}\c@page\z@
	\def\thefootnote{\fnsymbol{footnote}} }

\def\endtitlepage{\if@restonecol\twocolumn \else  \fi
	\def\thefootnote{\arabic{footnote}}
	\setcounter{footnote}{0}}  
\catcode`@=12
\relax

\def\beq{\begin{equation}}
\def\eeq{\end{equation}}
\def\bea{\begin{eqnarray}}
\def\eea{\end{eqnarray}}
\def\nn{\nonumber}

\relax
\hybrid

\begin{document}

\begin{titlepage}
\begin{center}
November~2019 \hfill . \\[.5in]
{\large\bf Four spins correlation function of the $q$ states Potts model,
  for general values of $q$. Its percolation model limit $q\rightarrow 1$. }
\\[.5in] 
{\bf Vladimir S.~Dotsenko}\\[.2in]
{\it LPTHE, 
Sorbonne Universit{\'e}, CNRS, UMR 7589\\
4 place Jussieu,75252 Paris Cedex 05, France.}\\[.2in]
               
 \end{center}
 
\underline{Abstract.}

Under the assumption that the product of two spin operators decomposes
uniquely into the degenerate conformal fields $\{\Phi_{n',n}\}$,
the general expression for the correlation function of four spins is defined
for the $q$ states Potts model with $q$ taking general values in the interval $1\leq q\leq 4$.

The limit of $q\rightarrow 1$ is considered in detail and the four spin function
is obtained for the percolation model.

\end{titlepage}

\newpage

\numberwithin{equation}{section}

\section{Introduction.}

The renewed interest in the critical Potts model is related to the studies
of three and four points cluster connectivities, which are connected
to the coorresponding spin correlation functions 
\cite{ref1a} - \cite{ref1f}.

In the context of $2D$ conformal field theory, of the $q$ states 
Potts model \cite{ref1,ref2}, the four spins function:
\beq
<\sigma(\infty)\sigma(1)\sigma(z,\bar{z})\sigma(0)> \label{eq1.1}
\eeq
is a complicated object. This is because the spin operator is not a degenerate field,
for general values of $q$, unlike for instance the energy operator
which is the $\Phi_{1,2}$ degenerate field, in the conformal field theory classification,
for general values of $q$ in the interval $1\leq q\leq 4$.

For the degenerate operators, the conformal field theory provides
well defined methods for calculation, in particular,
of four-point functions \cite{ref3,ref2,ref4}. For non-degenerate operators
such techniques are absent.

In the next Section we shall suggest the method which allows
to turn around this difficulty, in order to define the four spins function (\ref{eq1.1})
for general values of $q$.

To be more precise, this is done under the assumption,
yet to be justified, that the product of two spin operators decomposes
uniquely into the degenerate conformal fields $\{\Phi_{n',n}\}$. 
This assumption is presented in some more details further below, 
in the next Section, in the set of remarks which follow the equation (2.36).

In the third Section we shall consider 
the limit $q\rightarrow 1$ 
for the function (\ref{eq1.1}), which is the percolation model 
four spins function. This limit turns out to be very delicate.

The Section 4 is devoted to the discussion.

\section{General expression for the four spins correlation function of the $q$ states Potts model.}

The full space of degenerate operators of the conformal field theory with the central charge $c$ in the interval
\beq
0\leq c \leq 1 \label{eq2.1}
\eeq
is covered by the fields \cite{ref3}:
\beq
\{\Phi_{n',n}\} \label{eq2.2}
\eeq
$n'=1,2,3,...,\infty$, $n=1,2,3,...,\infty$, having conformal dimensions
\beq
\{\Delta_{n',n}\} \label{eq2.3}
\eeq
which are given by the Kac formula
\beq
\Delta_{n',n}
=\frac{(\alpha_{-}n'+\alpha_{+}n)^{2}-(\alpha_{-}+\alpha_{+})^{2}}{4}, 
\label{eq2.4}
\eeq
\beq
\alpha_{\pm}=\alpha_{0}\pm\sqrt{\alpha^{2}_{0}+1}, \label{eq2.5}
\eeq
\beq
c=1-24\alpha_{0}^{2} \label{eq2.6} 
\eeq
Above is used the Coulomb Gas (CG)  based parametrisation. $\alpha_{-}$, $\alpha_{+}$ are the charges of the screening operators, $\alpha_{0}$ is the background charge of the vacuum \cite{ref2}. The CG vertex operators representing the primary fields (\ref{eq2.2}) have the form:
\beq
V_{n',n}(z,\bar{z})\equiv V_{\alpha_{n',n}}(z,\bar{z})
=\exp\{i\alpha_{n',n}\varphi(z,\bar{z})\} \label{eq2.7}
\eeq
$\varphi(z,\bar{z})$ is a free scalar field and the corresponding CG charges $\{\alpha_{n',n}\}$ have the form:
\beq
\alpha_{n',n}
=\frac{1-n'}{2}\alpha_{-}+\frac{1-n}{2}\alpha_{+}
\label{eq2.8}
\eeq

If, in turn, $\alpha^{2}_{+}$ is parametrized as
\beq
\alpha_{+}^{2}=\frac{p+1}{p} \label{eq2.9}
\eeq
then, for the parameter $p$ taking odd integer values
\beq
p=3,5,7,...\infty \,,
\label{eq2.10}
\eeq
the corresponding conformal field theories, 
with the central charge
\beq
c_{p}=1-\frac{6}{p(p+1)}\,\, , \label{eq2.11}
\eeq
correspond to minimal models representing the unitary set of Potts models, with $q$ taking a discrete (infinite) set of values ranging from  $q=2\,\, (p=3,\, c=1/2)$
to $q=4\,\, (p\rightarrow\infty,\, c=1)$  \cite{ref5}. 
More precisely, $q$ is related to the parameter $p$
by the formula:
\beq
\sqrt{q} = 2 \cos\frac{\pi}{p+1} \label{eq2.12}
\eeq
More details could be found in \cite{ref1,ref2,ref5}.

To come to $c_{p}$ in (\ref{eq2.11}), from the central charge expression in (\ref{eq2.6}), the standard relations 
are to be used:
\bea
\alpha_{+}
=\sqrt{\frac{p+1}{p}},\,\,\,\, \alpha_{+}\alpha_{-}=-1,
\,\,\,\,\alpha_{-}=-\sqrt{\frac{p}{p+1}}\nn\\
\alpha_{0}=\frac{\alpha_{+}+\alpha_{-}}{2}=\frac{1}{2\sqrt{p(p+1)}} \label{eq2.13}
\eea

For a particular unitary minimal model, with 
\beq
p=2s+1\,\, \label{eq2.14}
\eeq
$s$ being positive integer, a finite set of primary fields
(\ref{eq2.2}) decouple, from the rest, 
by the operator algebra. They form a finite table (Kac table):
\beq
\{\Phi_{n',n}\},\,\,1\leq n'\leq 2s+1,\,\,1\leq n\leq 2s
\label{eq2.15}
\eeq
In fact, the number of primary fields, of the corresponding model, is twice less, because of the symmetry: the fields
\beq
\Phi_{n',n}\,\,\,\mbox{and}\,\,\, \Phi_{2s+2-n',2s+1-n}
\label{eq2.16}
\eeq
are identical. They have the same conformal dimensions and identical conformal theory properties.

The spin operator $\sigma$, of the $M_{p}$ minimal model (\ref{eq2.15}), finds itself in the middle
of the Kac table.
Its two copies are the operators: 
\beq
\Phi_{\frac{p+1}{2},\frac{p-1}{2}}\,\,\,\mbox{and}\,\,\,\Phi_{\frac{p+1}{2},\frac{p+1}{2}}
\label{eq2.17}
\eeq
or, for $p=2s+1$:
\beq
\Phi_{s+1,s}\,\,\,\mbox{and}\,\,\,\Phi_{s+1,s+1}
\label{eq2.18}
\eeq

Now, for general (real) values of the parameter $p$, in the interval $2\leq p<\infty$, and hence the general values of the central charge (\ref{eq2.11}), in the interval $0\leq c\leq 1$, the Kac table of the degenerate fields (2.2) is infinite, while the spin field, equations (\ref{eq2.17}), 
(\ref{eq2.18}), no longer belongs to the Kac table, 
its indices being real numbers, in general, 
instead of positive integers. Still the spin operator is represented by the two fields in (\ref{eq2.18}).
Their conformal dimensions are equal if $s=(p-1)/2$.

The dimensions of the fields in (\ref{eq2.17}) could still be calculated by the formula (\ref{eq2.4}). More explicitly, with the parametrization by the parameter $p$, this formula 
is of the form:
\beq
\Delta_{n',n}=\frac{(pn'-(p+1)n)^{2}-1}{4p(p+1)}
\label{eq2.19}
\eeq
For $n'=(p+1)/2$, $n=(p-1)/2$, of the first field 
in (\ref{eq2.17}), and also for $n'=(p+1)/2$, $n=(p+1)/2$, 
for the second field in (\ref{eq2.17}), 
one gets the same value:
\beq
\Delta_{\sigma}=\frac{(p-1)(p+3)}{16p(p+1)}
\label{eq2.20}
\eeq
-- the conformal dimension of the spin field of the Potts model, for general values of the parameter $p$.

The fact that the spin operator has two representations, 
as in (\ref{eq2.17}), for general values of $p$ 
(or $c$, or $q$), in the same way as it was the case
for the minimal models when $p$ was odd integer, 
this fact will be important in the following.

In the context of a particular minimal model $M_{p}$, 
with $p$ being odd integer, the parameter
\beq
s=\frac{p-1}{2} \label{eq2.21}
\eeq
in (\ref{eq2.18}) being positive integer, the correlation 
function of four spins (\ref{eq1.1}) could readily be calculated with the Coulomb Gas technique 
\cite{ref2,ref4}. But for general values of $p$ this way of calculating the four spin function is blocked: one would need to use a fractional number of screening operators, 
a fractional number of integrations, 
in the technique of \cite{ref2,ref4}.

On the other hand, the four-point function
\beq
<\sigma(\infty)\Phi_{n',n}(1)\Phi_{n',n}(z,\bar{z})\sigma(0)>
\label{eq2.22}
\eeq
could readily be calculated with the CG technique. It could be represented, in the CG technique, by the four-point function
\beq
<V_{\alpha^{+}_{\sigma}}(\infty)V_{\alpha_{n',n}}(1)V_{\alpha_{n',n}}(z,\bar{z})V_{\alpha_{\sigma}}(0)>_{conf}
\label{eq2.23}
\eeq
The CG charge of the operators $V_{\alpha_{n',n}}(1)$, $V_{\alpha_{n',n}}(0)$ is given by the formula (\ref{eq2.8}).
The charge of the spin operator $V_{\alpha_{\sigma}}(0)$ 
is also defined by the formula (\ref{eq2.8}), but with the fractional indices $n'=(p+1)/2$, $n=(p-1)/2$
in the case of the first operator in (\ref{eq2.17}):
\beq
\alpha_{\sigma}=\alpha_{\frac{p+1}{2},\frac{p-1}{2}}=\frac{1-\frac{p+1}{2}}{2}\alpha_{-}+\frac{1-\frac{p-1}{2}}{2}\alpha_{+}=\frac{1-p}{4}\alpha_{-}+\frac{3-p}{4}\alpha_{+}
\label{eq2.24}
\eeq
The charge of the operator $V_{\alpha^{+}_{\sigma}}(\infty)$, representing the spin operator at infinity, is CG conjugate:
\beq
\alpha^{+}_{\sigma}=2\alpha_{0}-\alpha_{\sigma}
\label{eq2.25}
\eeq
The total CG charge of the four vertex operators 
in (\ref{eq2.23}) is
\beq
2\alpha_{0}-\alpha_{\sigma}+2\alpha_{n',n}+\alpha_{\sigma}=2\alpha_{0}+2\alpha_{n',n}=2\alpha_{0}+(1-n')\alpha_{-}+(1-n)\alpha_{+} 
\label{eq2.26}
\eeq
According to the CG technique \cite{ref2,ref4}, 
this function requires $n'-1$ screening operators
\beq
\int d^{2}uV_{\alpha_{-}}(u,\bar{u}),
\quad\,\,\,   V_{\alpha_{-}}(u,\bar{u})
=e^{i\alpha_{-}\varphi(u,\bar{u})}
\label{eq2.27}
\eeq
and $n-1$ screening operators
\beq
\int d^{2}v\,V_{\alpha_{+}(v,\bar{v})},
\quad\,\,\,V_{\alpha_{+}}(v,\bar{v})
=e^{i\alpha_{+}\varphi(v,\bar{v})}
\label{eq2.28}
\eeq
This means that the four-point function (\ref{eq2.23}), 
and hence the function (\ref{eq2.22}), will be expressed by the multiple integral of \cite{ref4} and it could be calculated.

Now, in more detail, the function (\ref{eq2.22}) could be represented, in the $s$ - channel decomposition (decomposition in powers of $z,\bar{z}$), as follows:
\bea
<\sigma(\infty)\Phi_{n',n}(1)\Phi_{n',n}(z'\bar{z})\sigma(0)>\nn\\
=\sum_{t',t}\frac{1}{|z|^{2\Delta_{\sigma}+2\Delta_{n',n}-2\Delta_{t',t}}}\times (D^{(t',t)}_{\sigma,(n',n)})^{2}|F_{t',t}(z)|^{2}
\label{eq2.29}
\eea
$F_{t',t}(z)$ is the conformal block function of the intermediate channel $(t',t)$, which is defined, in the standard way, by the Virassoro algebra \cite{ref3}. This function depends also on the four external operators which dependence have been suppressed.

The operator algebra constants in (\ref{eq2.29})
\beq
\{D^{(t',t)}_{\sigma,(n',n)}
\equiv D^{\Phi_{t',t}}_{\sigma,\Phi_{n',n}}\} \label{eq2.30}
\eeq
-- they are given by the multiple integrals of \cite{ref4}. 
They are expressed finally by products of $\Gamma$ 
functions.

When $n'$ is given odd values, $n'=2l+1$, $l=0,1,2,3,...$,
\bea
<\sigma(\infty)\Phi_{2l+1,n}(1)\Phi_{2l+1,n}(z\bar{z})\sigma(0)>\nn\\
=\sum_{(t',t)}\frac{1}{|z|^{2\Delta_{\sigma}+2\Delta_{2l+1,n}-2\Delta_{t',t}}}(D_{\sigma,(2l+1,n)}^{(t',t)})^{2}|F_{t',t}(z)|^{2} \,,
\label{eq2.31}
\eea
in this case among the operators 
of the intermediate channels 
in (\ref{eq2.31}), $\{\Phi_{t',t}\}$, there is the spin operator 
$\sigma$: the channel of the first operator in (\ref{eq2.17}) 
if $n$ is odd, and the channel of the second operator in (\ref{eq2.17}) if $n$ is even, Fig.1. The operator algebra 
constant for this particular channel, 
\beq
D^{\sigma}_{\sigma,(2l+1,n)}\,\,,
\label{eq2.32}
\eeq
in the sum of (\ref{eq2.31}),
corresponds to fusion:
\beq
\sigma\times\Phi_{(2l+1,n)}\rightarrow \sigma
\label{eq2.33}
\eeq
But this process could be turned around as in Fig.2,
\beq
D^{\sigma}_{\sigma,(2l+1,n)}
=D_{\sigma,\sigma}^{(2l+1,n)}
\label{eq2.34}
\eeq

\begin{figure}
\begin{center}
\epsfxsize=300pt\epsfysize=170pt{\epsffile{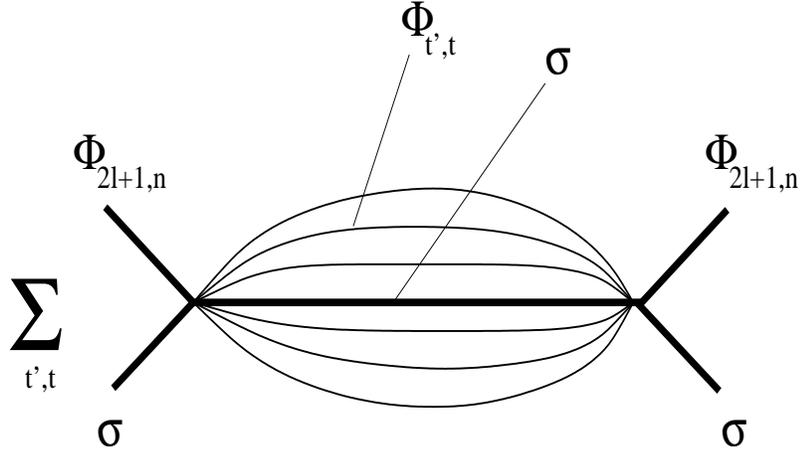}}
\caption{One intermediate channel, out of many, is that
of $\{\Phi_{t',t}\} = \sigma$.
\label{fig1}
}
\end{center}
\end{figure}

\begin{figure}
\begin{center}
\epsfxsize=300pt\epsfysize=120pt{\epsffile{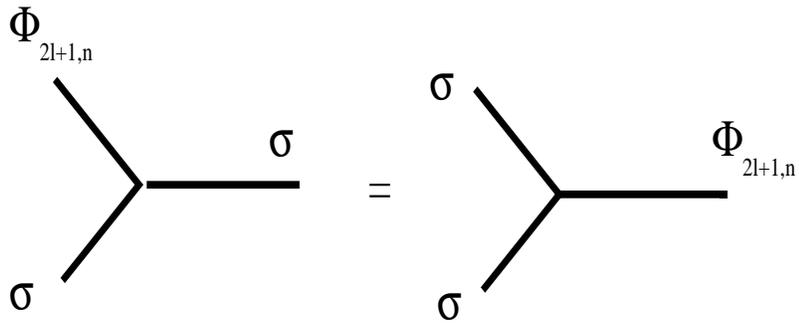}}
\caption{The symmetry of the operator algebra constants
$D^{\sigma}_{\sigma,(2l+1,n)}$, eq.(\ref{eq2.34}).
\label{fig2}
}
\end{center}
\end{figure}

We remind that for the normalised operators, 
the operator algebra constants are symmetric 
in all three indices, i.e. in all three operators. 
In this way, from one particular intermediate 
channel of the four point function in (\ref{eq2.31}), 
we get the constants for the product 
of two spin operators producing the fields
\beq
\{\Phi_{2l+1,n}\} \label{eq2.35}
\eeq
In other words, we get the operator algebra constants for the intermediate channels of the function
\bea
<\sigma(\infty)\sigma(1)\sigma(z,\bar{z})\sigma(0)>\nn\\
=\sum_{(l,n)}\frac{1}{|z|^{4\Delta_{\sigma}-2\Delta_{2l+1,n}}}(D^{(2l+1,n)}_{\sigma,\sigma})^{2}|F_{(2l+1,n)}(z)|^{2}
\label{eq2.36}
\eea

Stating it again, the constants in the decomposition 
(\ref{eq2.36}) above, by using the symmetry 
(\ref{eq2.34}), Fig.2, 
could readily be calculated by the well defined 
integrals which define the four-point function (\ref{eq2.31}). 
This is done by extracting an appropriate 
channel from it, from the sum in (\ref{eq2.31}), Fig1.

Several remarks could be added at this point.

The three-point functions could be defined for any three operators \cite{ref6,ref1c,ref8}. 
In general, they correspond to projecting a couple of operators on the third one. They do not, in general, represent the operator algebra constants of a particular theory, the constants which would appear, naturally, in the decomposition of four-point functions.

The constants $D^{\sigma}_{\sigma,(2l+1,n)}$, eq.(\ref{eq2.32}),
which could also be calculated as three-point functions, they do represent the operator algebra constants, of the Potts model, as they originate in the four-point function 
(\ref{eq2.31}). They could be called physical in the above sense. In this case the constants 
$D^{(2l+1,n)}_{\sigma,\sigma}$, in (\ref{eq2.34}), 
are also physical, they correspond to real processes in the model. They define, in particular, the four point function $<\sigma\sigma\sigma\sigma>$, as in eq.(\ref{eq2.36}).

As has been stated above with respect to the four-point function (\ref{eq2.29}), the spin operator appears 
as one of the intermediate channels of this function 
in the case when the index $n'$ is an odd integer. 
For $n'$ even, the spin operator is not amont the intermediate channel operators. Which means, in turn, 
that the primary operators $\{\Phi_{n',n}\}$, 
with $n'$ even, do not appear in the fusion of two spin operators. The corresponding operator algebra 
constants are zero.
In this case the sum in (\ref{eq2.36}) could be considered as being taken over all the primary fields,
of the Potts model, in the intermediate channels. 

We consider, but this is clearly an assumption,
that the set of primary fields $\{\Phi_{n',n}\}$ is complete
in the "neutral", or "even" sector:
the sector which is generated by fusing an even number of spins,
for the model with $q$ general, a real number.

The spin type operators, or "odd" sector fields,
are generated by fusing the spin
operator with the fields $\{\Phi_{n',n}\}$, 
with the energy operator $\Phi_{1,2}$ to begin with.
The fusion rule is clear from the Coulomb Gas integral representation
of the function $<\sigma(\infty)\Phi_{n',n}(1)\Phi_{n',n}(z,\bar{z})\sigma(0)>$,
eq.(2.29).
The sum over $(t',t)$, eq.(2.29), is being taken over fractional values:
the fractional indices of the operator $\sigma$, eq.(2.17),
being shifted by integer values.

In other words, the fusion rule
$\sigma\times\Phi_{n',n}\rightarrow\sum_{t',t}\Phi_{t',t}$
is the same as that of minimal models. The indices
of the operators $\{\Phi_{t',t}\}$,
which we call spin type operators, have fractional values which are the same
as those of the spin operator but shifted by integer values.
For instance,
$ \sigma\times\Phi_{1,2}\equiv \Phi_{\frac{p+1}{2},\frac{p+1}{2}}\times\Phi_{1,2}
\rightarrow \Phi_{\frac{p+1}{2},\frac{p-1}{2}}+\Phi_{\frac{p+1}{2},\frac{p+3}{2}}$.
$ \Phi_{\frac{p+1}{2},\frac{p-1}{2}} $ is the spin operator, $\sigma $,
and $\Phi_{\frac{p+1}{2},\frac{p+3}{2}}$ is a particular spin type operator.
The conformal dimensions of non-degenerate spin type operators
$\{\Phi_{t',t}\}$ are still given by the formula (2.19), as follows
from the Coulomb Gas. The operator algebra coefficients for fusions 
$\sigma\times\Phi_{n',n}\rightarrow\sum_{t',t}\Phi_{t',t}$ could all be calculated
by the corresponding  Coulomb Gas integrals, the way this is done
for a particular fusion channel, $\sigma\times\Phi_{n',n}\rightarrow \sigma $,
in the Section 3.

Under the assumption made above, on the completeness of the set
of degenerate fields $\{\Phi_{n',n}\}$ for the even sector,
the decomposition (\ref{eq2.36}),
for the function $<\sigma\sigma\sigma\sigma>$, 
will be exact. It defines the four-spin function of the Potts model.

We shall finish this Section by giving formulas for the operator
algebra coefficients $(D_{\sigma,\sigma}^{(2l+1,n)})^{2}$ in (\ref{eq2.36}).

As has been stated above, they are defined by a particular channel in the $s$-channel decomposition (\ref{eq2.31}), for the four-point function $<\sigma(\infty)\Phi_{2l+1,n}(1)\Phi_{2l+1,n}(z,\bar{z})\sigma(0)>$. This function is well defined by the Coulomb Gas integrals:
\bea
<\sigma(\infty)\Phi_{2l+1,n}(1)\Phi_{2l+1,n}(z,\bar{z})\sigma(0)>\nn\\
\propto<V_{\alpha^{+}_{\sigma}}(\infty)V_{\alpha_{2l+1,n}}(1)V_{\alpha_{2l+1,n}}(z,\bar{z})V_{\alpha_{\sigma}}(0)>_{conf}
\label{eq2.37}
\eea
The cases of values of the index $n$ being odd and even have to be considered separately.

For $n$ taking odd values, $n=2k+1$, the total Coulomb Gas charge in the function (\ref{eq2.37}) is given by:
\bea
\alpha^{+}_{\sigma}+2\alpha_{2l+1,2k+1}+\alpha_{\sigma}\nn\\
=2\alpha_{0}-\alpha_{\sigma}+(-2l\alpha_{-}-2k\alpha_{+})
+\alpha_{\sigma}\nn\\
=2\alpha_{0}-2l\alpha_{-}-2k\alpha_{+}
\label{eq2.38}
\eea
Accordingly, the Coulomb Gas function (\ref{eq2.37}) 
$(n=2k+1)$ requires $2l$ screenings $\alpha_{-}$ 
and $2k$ screening $\alpha_{+}$. One obtains: 
\bea
<V_{\alpha^{+}_{\sigma}}(\infty)V_{\alpha_{2l+1,2k+1}}(1)V_{\alpha_{2l+1,2k+1}}(z,\bar{z})V_{\alpha_{\sigma}}(0)>_{conf}\nn\\
=\prod^{2l}_{i=1}\int d^{2}u_{i}\prod^{2k}_{j=1}\int d^{2}v_{j}
<V_{\alpha^{+}_{\sigma}}(\infty)V_{\alpha_{2l+1,2k+1}}(1)V_{\alpha_{2l+1,2k+1}}(z,\bar{z})V_{\alpha_{\sigma}}(0)\nn\\
\prod^{2l}_{i=1}V_{\alpha_{-}}(u_{i},\bar{u}_{i})\prod_{j=1}^{2k}V_{\alpha_{+}}(v_{j},\bar{v}_{j})>_{(2\alpha_{0})}
\label{eq2.39}
\eea
The indexing $<...>_{conf}$, $<...>_{(2\alpha_{0})}$
should be clear from the equations above, for those
familiar with the CG technique.

Different channels in (\ref{eq2.31}) correspond to different separations of screenings: in one subset, the screenings 
are integrated around $0$ and $(z,\bar{z})$ 
while the rest of screenings are integrated 
around $1$ and $\infty$.
In the limit of $z\rightarrow 0$ this separation is exact 
and, accordingly, the coefficients 
$(D_{\sigma,(2l+1,2k+1)}^{(t',t)})^{2}$ 
are being equal to the product of two 3-point functions.

The particular channel, when $\Phi_{(t',t)}=\sigma$, is obtained when $l$ screenings $\alpha_{-}$ and $k$ screenings $\alpha_{+}$ are being integrated around $0$ and $(z,\bar{z})$, while the remaining $l$ and $k$ screenings are integrated around $1$ and $\infty$: 
over the whole plane in fact, neglecting the presence of operators at $0$ and $(z,\bar{z})$ and 
the other screenings running around them.

Identification of operators in the intermediate 
channels of (\ref{eq2.31}), of the spin operator in particular, is obtained by calculating the total Coulomb Gas charge of two vertex operators in the product of (\ref{eq2.39}), 
at $0$ and $(z,\bar{z})$ for instance, and the cloud of screenings running around them. In particular, when the operators at $0$ and $(z,\bar{z})$, $V_{\alpha_{\sigma}}(0)$ and $V_{\alpha_{2l+1,2k+1}}(z,\bar{z})$, are surrounded by $l$ and $k$ screenings, the total charge of this subset of operators is given by:
\bea
\alpha_{\sigma}+\alpha_{2l+1,2k+1}+l\alpha_{-}+k\alpha_{+}\nn\\
=\alpha_{\sigma}+(-l\alpha_{-}-k\alpha_{+})+l\alpha_{-}+k\alpha_{+}=\alpha_{\sigma}
\label{eq2.40}
\eea
So that the corresponding intermediate channel 
operator is in fact the spin operator.

In correspondance to the above described factorisation of the total set of operators (in the limit of $z\rightarrow 0$), into a subset of operators located around $0$ and 
$(z,\bar{z})$ and an another subset located around $1$ and $\infty$, the coefficients, squared, $(D_{\sigma,(2l+1,2k+1)})^{2}$ are being expressed as a product 
of two 3-point functions:
\bea
(D_{\sigma,(2l+1,2k+1)}^{(t',t)})^{2}\nn\\
=<\sigma(\infty)\Phi_{2l+1,2k+1}(1)\Phi_{(t',t)}(0)><\Phi_{(t',t)}(\infty)\Phi_{2l+1,2k+1}(1)\sigma(0)>\nn\\
\propto<V_{\alpha_{\sigma}^{+}}(\infty)V_{2l+1,2k+1}(1)V_{\alpha_{t',t}}(0)>_{conf}<V_{\alpha^{+}_{t',t}}(\infty)V_{2l+1,2k+1}(1)V_{\alpha_{\sigma}}(0)>_{conf}
\label{eq2.41}
\eea
$V_{2l+1,2k+1}\equiv V_{\alpha_{2l+1,2k+1}}$. 
The variables $z,\bar{z}$ of $V_{2l+1,2k+1}(z,\bar{z})$
in the second 3-point function are assumed 
to have  been factored out by scaling.

In particular, for the spin operator 
in the intermediate channel:
\bea
(D_{\sigma,(2l+1,2k+1)}^{\sigma})^{2}\nn\\
=<\sigma(\infty)\Phi_{2l+1,2k+1}\sigma(0)><\sigma(\infty)\Phi_{2l+1,2k+1}(1)\sigma(0)>\nn\\
\propto<V_{\alpha_{\sigma}^{+}}(\infty)V_{2l+1,2k+1}(1)V_{\alpha_{\sigma}}(0)>_{conf}<V_{\alpha^{+}_{\sigma}}(\infty)V_{2l+1,2k+1}(1)V_{\alpha_{\sigma}}(0)>_{conf}\nn\\
=C_{\alpha_{\sigma}^{+},\alpha_{2l+1,2k+1},\alpha_{\sigma}}\times C_{\alpha_{\sigma}^{+},\alpha_{2l+1,2k+1},\alpha_{\sigma}}
\label{eq2.42}
\eea

The $C$ constants are the operator algebra coefficients of the Coulomb Gas vertex operators.

The proportionality in (\ref{eq2.42}) becomes the equality, when normalisation of the Coulomb Gas operators is taken into account \cite{ref8}. One gets:
\bea
<V_{\alpha_{\sigma}^{+}}(\infty)V_{2l+1,2k+1}(1)V_{\alpha_{\sigma}}(0)><V_{\alpha^{+}_{\sigma}}(\infty)V_{2l+1,2k+1}(1)V_{\alpha_{\sigma}}(0)>\nn\\
=<V_{\alpha_{\sigma}^{+}}(\infty)V_{2l+1,2k+1}(1)V_{\alpha_{\sigma}}(0)>^{2}\nn\\
=N(V_{\alpha^{+}_{\sigma}})^{2}\times N(V_{2l+1,2k+1})^{2}\times (N(V_{\alpha_{\sigma}}))^{2}\times <\sigma(\infty)\Phi_{2l+1,2k+1}(1)\sigma(0)>^{2}\nn\\
=\frac{1}{N(V_{\alpha_{\sigma}})^{2}\cdot Z^{2}}\times N^{2}_{2l+1,2k+1}\times N(V_{\alpha_{\sigma}})^{2}\times(D^{\sigma}_{\sigma,(2l+1,2k+1)})^{2}\nn\\
=\frac{1}{Z^{2}}N^{2}_{2l+1,2k+1}(D^{\sigma}_{\sigma,(2l+1,2k+1)})^{2}
\label{eq2.43}
\eea
Finally
\beq
(D^{\sigma}_{\sigma,(2l+1,2k+1)})^{2}=\frac{Z^{2}}{N^{2}_{2l+1,2k+1}}(C_{\alpha^{+}_{\sigma},\alpha_{2l+1,2k+1,\alpha_{\sigma}}})^{2}
\label{eq2.44}
\eeq
Here $Z$ is the Coulomb Gas partition function \cite{ref8}:
\beq
Z=\frac{-\rho}{(1-\rho)^{2}}\gamma(\rho')\gamma(\rho),
\quad\,\rho'=\alpha^{2}_{-},\quad\,\rho=\alpha^{2}_{+},
\quad\,\gamma(x)=\Gamma(x)/\Gamma(1-x)
\label{eq2.45}
\eeq

$N_{2l+1,2k+1}$ is the norm of the operator $V_{2l+1,2k+1}\equiv V_{\alpha_{2l+1,2k+1}}$. Is used also the relation between the norms of the operators $V^{+}_{\alpha}$ and $V_{\alpha}$ \cite{ref8}:
\beq
N(V_{\alpha^{+}})\equiv N(V_{\alpha}^{+})=\frac{1}{ZN(V_{\alpha})}
\label{eq2.46}
\eeq
The Coulomb Gas structure constants are given by 
\cite{ref4,ref8}:
\bea
C_{c,b,a}=<V_{c}(\infty)V_{b}(1)V_{a}(0)>_{conf}\nn\\
=\frac{\rho^{-4lk}}{Z}\times\prod^{l}_{i=1}\gamma(i\rho'-k)\times\prod^{k}_{j=1}\gamma(j\rho)\nn\\
\times\prod^{l-1}_{i=0}\gamma(1-k+\alpha'+i\rho')\gamma(1-k+\beta'+i\rho')\gamma(1-k+\gamma'+i\rho')\nn\\
\times\prod^{k-1}_{j=0}\gamma(1+\alpha+j\rho)\gamma(1+\beta+j\rho)\gamma(1+\gamma+j\rho)\nn\\
=\frac{\rho^{-4lk}}{Z}\prod^{l}_{i=1}\prod^{k}_{j=1}\frac{1}{(i\rho'-j)^{2}}\times\prod^{l}_{i=1}\gamma(i\rho')\prod^{k}_{j=1}\gamma(j\rho)\nn\\
\times\prod^{l-1}_{i=0}\prod^{k-1}_{j=0}\frac{1}{(\alpha'+i\rho'-j)^{2}(\beta'+i\rho'-j)^{2}(\gamma'+i\rho'-j)^{2}}\nn\\
\times\prod^{l-1}_{i=0}\gamma(1+\alpha'+i\rho')\gamma(1+\beta'+i\rho')\gamma(1+\gamma'+i\rho')\nn\\
\times\prod^{k-1}_{j=0}\gamma(1+\alpha+j\rho)\gamma(1+\beta+j\rho)\gamma(1+\gamma+j\rho)
\label{eq2.47}
\eea
-- equations (4.4),(4.8) of \cite{ref8}, 
with the normalization factor $1/Z$ added. In the equations above $l$ and $k$ are the numbers of screening operators needed by the Coulomb Gas correlator $<V_{c}(\infty)V_{b}(1)V_{a}(0)>_{conf}$.
$a,b,c$ are the Coulomb Gas charges of the operators $V_{a},V_{b},V_{c}$.
The parameters $\alpha',\beta',\gamma',\alpha, \beta,\gamma$ are given by:
\bea
\alpha'=2\alpha_{-}a,\,\,\beta'=2\alpha_{-}b,\,\,\gamma'=2\alpha_{-}c,\nn\\
\alpha=2\alpha_{+}a,\,\,\beta=2\alpha_{+}b,\,\,\gamma=2\alpha_{+}c
\label{eq2.48}
\eea

Our excuses for using $\alpha$ (with indexes) 
for the Coulomb Gas charges is the previous equations, while also as a parameter (without indexes) 
in the equations (\ref{eq2.47}), (\ref{eq2.48}).

Similarly, our excuses for using $\gamma$ for the ratio
of $\Gamma$ functions ($\gamma(x)$, eq.(\ref{eq2.45}))
in the equation (\ref{eq2.47}), while also 
for the parameters $\gamma,\,\,\gamma'$ 
in (\ref{eq2.47}), (\ref{eq2.48}).

The norm squared $N^{2}_{2l+1,2k+1}$ in the equation (\ref{eq2.44}), of the operator $V_{2l+1,2k+1}\equiv V_{\alpha_{2l+1,2k+1}}$, is given by the general  
equation \cite{ref8}:
\bea
N(V_{\alpha_{n',n} })=<I^{+}(\infty)V_{\alpha_{n',n}}(1)V_{\alpha_{n',n}}(0)>\nn\\
=\prod^{n'-1}_{i=1}\frac{\gamma(-1+(1+i)\rho')}{\gamma(i\rho')}\times\prod^{n-1}_{j=1}\frac{\gamma(-1+(1+j)\rho)}{\gamma(j\rho)}\nn\\
\times\prod^{n'-1}_{i=1}\prod^{n-1}_{j=1}(-i+j\rho)^{2}\times\prod^{n'}_{i=2}\prod^{n}_{j=2}\frac{1}{(-i+j\rho)^{2}}
\label{eq2.49}
\eea
$I^{+}\equiv V_{2\alpha_{0}}$ is the CG conjugate identity operator.

The equation (\ref{eq2.44}) defines the structure constants $(D_{\sigma,(2l+1,n)}^{\sigma})^{2}$, and hence the constants $(D_{\sigma,\sigma}^{(2l+1,n)})^{2}$ in (\ref{eq2.36}), in the case when the index $n$ is an odd integer, $n=2k+1$. In the case of $n$ being an even integer, $n=2k$, the expressions are differences. In this case the Coulomb Gas function (\ref{eq2.37}),
\beq
<V_{\alpha_{\sigma}^{+}}(\infty)V_{2l+1,2k}(1)V_{2l+1,2k}(z,\bar{z})V_{\alpha_{\sigma}}(0)>_{conf}
\label{eq2.50}
\eeq
will require $2l$ screenings $\alpha_{-}$ and $2k-1$ screenings $\alpha_{+}$.

The spin operator appears in the intermediate $s$ channel of the function (\ref{eq2.37}), 
decomposed as in the equation (\ref{eq2.31}), 
when $l$ and $k-1$ screenings are integrated around 
$0$ and $(z,\bar{z})$, while $l$ and $k$ screenings 
are integrated around $1$ and $\infty$. 
In fact, in this case the total charge of the operators 
in the region around $0$ and $(z,\bar{z})$ is given by:
\bea
\alpha_{\sigma}+\alpha_{2l+1,2k}+l\alpha_{-}+(k-1)\alpha_{+}\nn\\
=\alpha_{\sigma}-l\alpha_{-}-(k-\frac{1}{2})\alpha_{+}+l\alpha_{-}+(k-1)\alpha_{+}\nn\\
=\alpha_{\sigma}-\frac{1}{2}\alpha_{+}
\label{eq2.51}
\eea
We remind that the spin operator, the first operator in (\ref{eq2.17}),
\beq
\sigma=\Phi_{\frac{p+1}{2},\frac{p-1}{2}}\propto V_{\alpha_{\sigma}}
\label{eq2.52}
\eeq
has the CG charge:
\beq
\alpha_{\sigma}=\alpha_{\frac{p+1}{2},\frac{p-1}{2}}=\frac{1-p}{4}\alpha_{-}+\frac{3-p}{4}\alpha_{+}
\label{eq2.53}
\eeq
Then, by (\ref{eq2.51}), the total charge 
in the $0$, $(z,\bar{z})$ region is equal to:
\beq
\alpha_{\sigma}-\frac{1}{2}\alpha_{+}
=\frac{1-p}{4}\alpha_{-}+\frac{1-p}{4}\alpha_{+}
=\alpha_{\frac{1+p}{2},\frac{1+p}{2}}
\label{eq2.54}
\eeq
which is the charge of the second operator in (\ref{eq2.17}), the second copy of the spin operator.

More precisely
\beq
\alpha_{\frac{p+1}{2},\frac{p+1}{2}}=\frac{1-p}{4}\alpha_{-}+\frac{1-p}{4}\alpha{+}=2\alpha_{0}-\alpha_{\sigma}=\alpha_{\sigma}^{+}
\label{eq2.55}
\eeq 
This equality is easily verified. On one hand,
\bea
\alpha_{\frac{p+1}{2},\frac{p+1}{2}}=\frac{1-p}{4}(\alpha_{-}+\alpha_{+}),\nn\\
\alpha_{-}=-\sqrt{\frac{p}{p+1}},\,\,\alpha_{+}=\sqrt{\frac{p+1}{p}},\nn\\
\alpha_{-}+\alpha_{+}=\frac{1}{\sqrt{p(p+1)}},\nn\\
\alpha_{\frac{p+1}{2},\frac{p+1}{2}}=\frac{1-p}{4}\cdot\frac{1}{\sqrt{p(p+1)}}
\label{eq2.56}
\eea
On the other hand
\bea
\alpha_{\sigma}^{+}=2\alpha_{0}-\alpha_{\sigma}=\alpha_{-}+\alpha_{-}-\alpha_{\sigma}\nn\\
=\frac{1}{\sqrt{p(p+1}}-\frac{1-p}{4}\alpha_{-}-\frac{3-p}{4}\alpha_{+}\nn\\
=\frac{1}{\sqrt{p(p+1)}}+\frac{1-p}{4}\cdot\sqrt{\frac{p}{p+1}}-\frac{3-p}{4}\sqrt{\frac{p+1}{p}}\nn\\
=\frac{1}{4\sqrt{p(p+1)}}[4+(1-p)p-(3-p)(p+1)]\nn\\
=\frac{1}{4\sqrt{p(p+1}}(1-p)
\label{eq2.57}
\eea
By (\ref{eq2.56}) and (\ref{eq2.57}), $\alpha_{\frac{p+1}{2},\frac{p+1}{2}}=\alpha^{+}_{\sigma}$, 
as stated in (\ref{eq2.55}).

Returning back to (\ref{eq2.54}), one finds, 
by the total CG charge of the operators 
in the region $0$, $(z,\bar{z})$, that the corresponding intermediate channel CG operator is
\beq
V_{\alpha^{+}_{\sigma}}\equiv V^{+}_{\alpha_{\sigma}}
\label{eq2.58}
\eeq
When projected onto the operator $V_{\alpha_{\sigma}}(\infty)$, to form the corresponding 3-point function, 
one obtains the contribution of the $0,(z,\bar{z})$ region, which factors out from the rest of the operators, 
to be given by:
\beq
<V_{\alpha_{\sigma}}(\infty)V_{2l+1,2k}(1)V_{\alpha_{\sigma}}(0)>=C_{\alpha_{\sigma},\alpha_{2l+1,2k},\alpha_{\sigma}}
\label{eq2.59}
\eeq
The variables $z$ and $\bar{z}$ of the operator $V_{2l+1,2k}(z,\bar{z})$, initially, 
are assumed to have been factored out by scaling.

In a similar way as is describe above, one checks that the rest of the operators in (\ref{eq2.50}), the operators $V_{\alpha_{\sigma}^{+}}(\infty)$ and $V_{2l+1,2k}(1)$ and $l$ and $k$ screenings integrated around, 
that this subset of operators correspond to the intermediate channel CG operator $V_{\alpha_{\sigma}}$. When projected onto the operator 
$V_{\alpha_{\sigma}^{+}}(0)$, to form 
the corresponding 3-point function, one finds that the contribution of the region $\infty,1$ which also factors 
out (in the limit $z\rightarrow 0$), is given by:
\beq
<V_{\alpha_{\sigma}^{+}}(\infty)V_{2l+1,2k}(1)V_{\alpha_{\sigma}^{+}}(0)>=C_{\alpha^{+}_{\sigma},\alpha_{2l+1,2k},\alpha^{+}_{\sigma}}
\label{eq2.60}
\eeq

Now, in the present case of the index $n$, 
of $(D^{\sigma}_{\sigma,(2l+1,n)})^{2}$, 
taking even values, the expression (\ref{eq2.44}) 
is replaced by the following one:
\beq
(D^{\sigma}_{\sigma,(2l+1,2k)})^{2}=\frac{Z^{2}}{N^{2}_{2l+1,2k}}\times
C_{\alpha^{+}_{\sigma},\alpha_{2l+1,2k},\alpha^{+}_{\sigma}}\times
C_{\alpha_{\sigma},\alpha_{2l+1,2k},\alpha_{\sigma}}
\label{eq2.61}
\eeq
$Z$ is given by the equation (\ref{eq2.45}), 
the normalisation constant squared $N^{2}_{2l+1,2k}$ 
is given by the general expression (\ref{eq2.49}) 
and the Coulomb Gas constants in (\ref{eq2.61}) 
are defined by the general formula (\ref{eq2.47}).

In this way, by the equations (\ref{eq2.44}), 
(\ref{eq2.61}), the structure constants $(D^{\sigma}_{\sigma,(2l+1,n)})^{2}$ are defined for all values of $n$, and hence the structure constants $D_{\sigma,\sigma}^{(2l+1,n)}$ in (\ref{eq2.36}), the $s$ channel 
decomposition of the four spin function $<\sigma(\infty)\sigma(1)\sigma(z,\bar{z})\sigma(0)>$.

It could be remarked that in the present case the expressions for the structure constants 
$(D_{\sigma,\sigma}^{(2l+1,n)})^{2}$, given 
in terms of CG constants 
expressed by products 
of the usual $\Gamma$ functions, 
are by far much simpler to analyse 
and calculate as compared to more general 
expressions for the 3-point functions 
in \cite{ref6,ref1c,ref8}, expressed in terms 
of $\Upsilon$ function.  

It could also be remarked that the general formula 
(\ref{eq2.36}), for the four spins function, might not be so straightforward to use in certain particular cases, for particular values of the parameter $p$. 
In particular, the reductions to a finite number of terms, in (\ref{eq2.36}), in cases of minimal models, have to be analysed carefully, as certain decouplings are not explicite, not simply given by the corresponding $D$ constants being zero, see the remarks in \cite{ref6} and \cite{ref8}.

One particular application of the general formula (\ref{eq2.36}) will be considered in the next Section, 
the percolation model limit of the four spins 
function (\ref{eq2.36}). The limit of $p\rightarrow 2$, $c\rightarrow 0$, or of
\beq
\alpha^{2}_{+}=\frac{3}{2}-\epsilon,
\,\quad\epsilon\rightarrow 0 \,,
\label{eq2.62}
\eeq
this limit turns out to be fairly complicated.

\section{The percolation model limit $q \rightarrow 1.$}

The four spins correlation function in the percolation model has already been considered, in particular, in \cite{ref9}. The limiting procedure for the spin operators, used in the paper \cite{ref9}, was somewhat forced, with the result that just one intermediate channel was selected, out of many, and not the most singular one in the limit $q\rightarrow 1$.
In a sense, in this paper we shall complete the study of the four spins function which was started in \cite{ref9}.

Our problem will be to take, properly, the limit $q\rightarrow 0$ for the expression of the four spins function in (\ref{eq2.36}).
To take the limit we shall use the parameter $\alpha^{2}_{+}\equiv\rho$ taken in the form
\beq
\alpha^{2}_{+}\equiv\rho=\frac{3}{2}-\epsilon
\label{eq3.1}
\eeq
The value $\alpha^{2}_{+}=3/2$ corresponds to the $q=1$, $c=0$ point, so that the limit $q\rightarrow 1$, $c\rightarrow 1$ corresponds to taking the limit $\epsilon\rightarrow 0$ in (\ref{eq3.1}). The operator algebra structure constants $(D^{(2l+1,n)}_{\sigma,\sigma})^{2}$ in (\ref{eq2.36}) have different limits, when $\epsilon\rightarrow 0$, dependng on values of $l$ and $n$. For some of them the limit is regular, they take finite or zero values when $\epsilon\rightarrow 0$. For others the limit is singular, they diverge 
as $1/\epsilon$ or  $1/\epsilon^{2}$.

To summerize the possibilities, 
when $\epsilon\rightarrow 0$,
\beq
(D^{(2l+1,n)}_{\sigma,\sigma})^{2}\rightarrow
\left\{\begin{array}{c}
\mbox{finite value or zero} \\
\sim 1/\epsilon\\
\sim 1/\epsilon^{2}
\end{array}\right.
\label{eq3.2}
\eeq
depending on values of $l$ and $n$. The expressions for the coefficients $(D^{(2l+1,n)}_{\sigma,\sigma})^{2}$ are given by (\ref{eq2.44}), for $n$ odd, $n=2k+1$, 
and by (\ref{eq2.61}) for $n$ even, $n=2k$. The tables 
in Fig.3 and Fig.4 summerize different limits for these coefficients, for different values of $l$ and $n$, odd and even. Most of the calculations have been done with the use of Mathematica.

\begin{figure}
\begin{center}
\epsfxsize=440pt\epsfysize=400pt{\epsffile{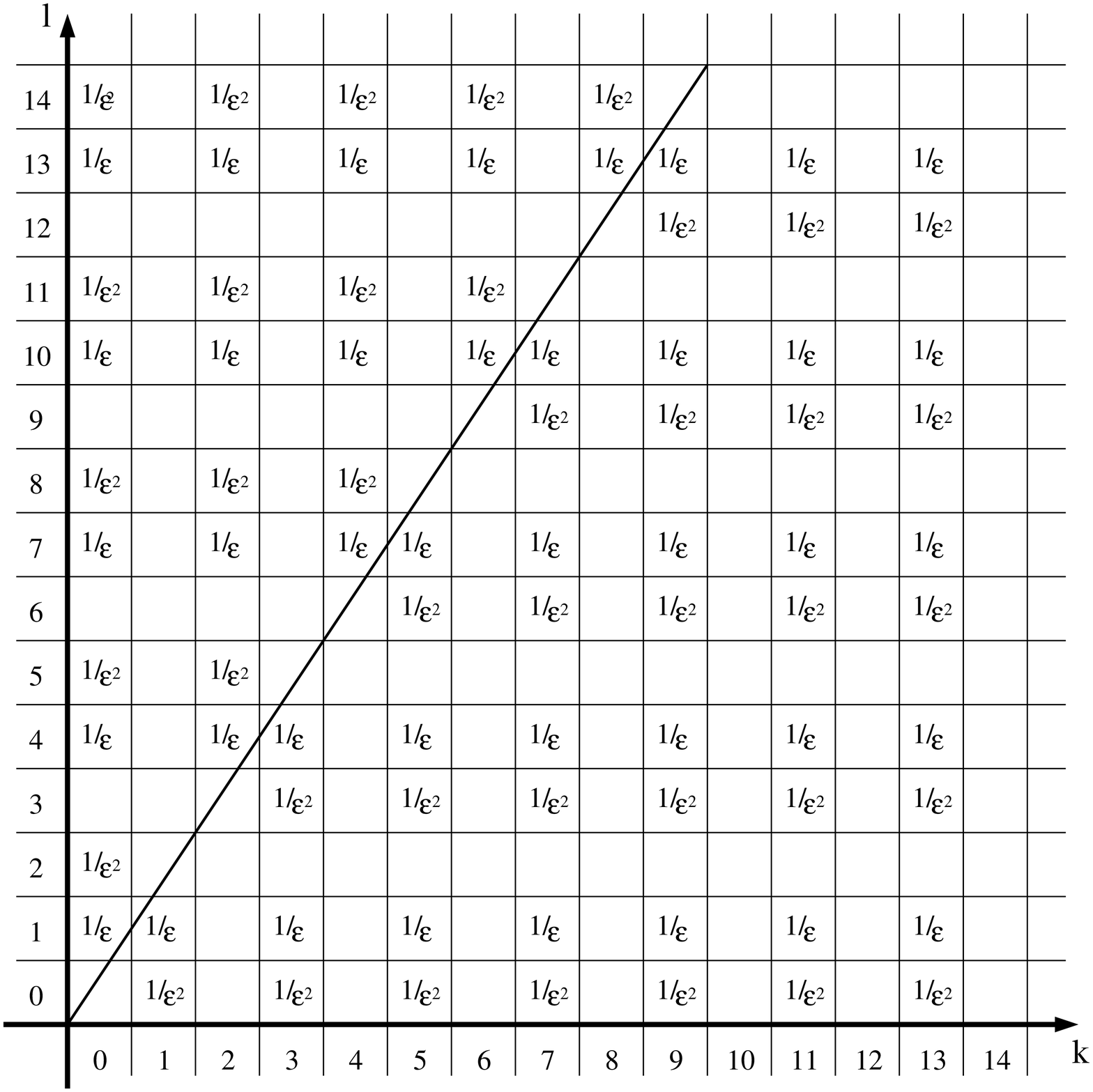}}
\caption{Different limits of the coefficients
$(D^{(2l+1,2k+1)}_{\sigma,\sigma})^{2}$ for different 
values of $l$ and $k$. 
Empty cells correspond to finite or zero limiting values.
\label{fig3}
}
\end{center}
\end{figure}

\begin{figure}
\begin{center}
\epsfxsize=440pt\epsfysize=400pt{\epsffile{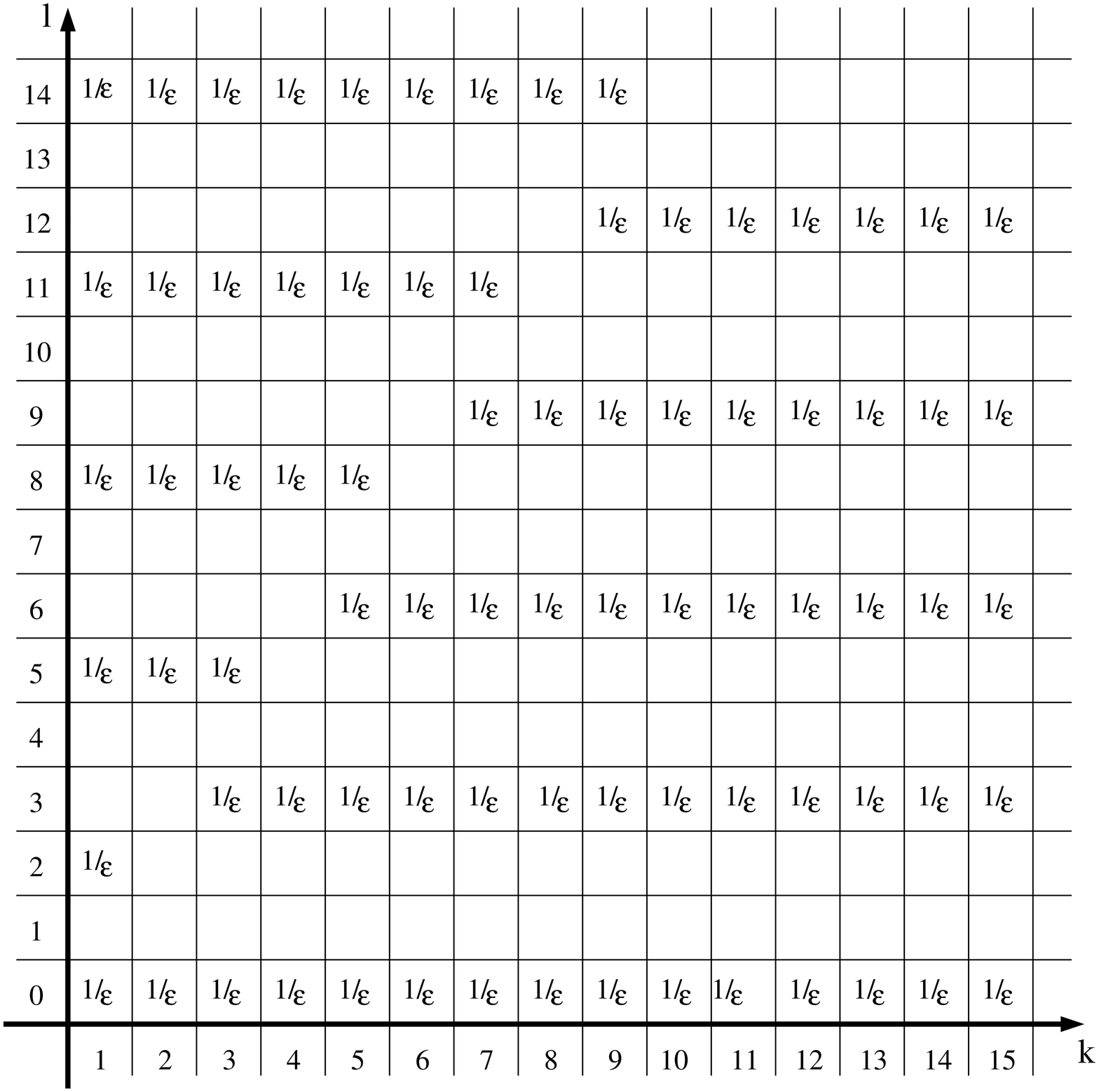}}
\caption{Different limits of the coefficients
$(D^{(2l+1,2k)}_{\sigma,\sigma})^{2}$ for different 
values of $l$ and $k$.
\label{fig4}
}
\end{center}
\end{figure}

Since the principal divergence is that of $1/\epsilon^{2}$, we shall define as the four spins function of the percolation model the limit:
\beq
\lim_{\epsilon\rightarrow 0}\epsilon^{2}<\sigma(\infty)\sigma(1)\sigma(z,\bar{z})\sigma(0)>
\label{eq3.3}
\eeq
In this limit only the $1/\epsilon^{2}$ intermediate channels will contribute, the rest will be suppressed.

Although, for $\epsilon$ small but different from zero, the other channels will perfectly be present. In particular, 
the channel of the energy operator $\Phi_{1,2}$, which is the cell of $l=0$, $k=1$ in the Fig.4.
This channel has been selected in \cite{ref9}, it diverges as $1/\epsilon$.

Still in this paper we shall study the four spins function as defined by the limit (\ref{eq3.3}). In this limit an infinity of intermediate channels will contribute, those which correspond to the diagonals $1/\epsilon^{2}$ in Fig.3.

The first diagonal is the one which starts with the cell 
$(l=0$, $k=1)$, the field $\Phi_{1,3}$, having 
the dimension
\beq
\Delta_{1,3}=\Delta(\Phi_{1,3})=2,
\label{eq3.4}
\eeq
in the limit $\epsilon\rightarrow 0$.

It is easily checked that all the fields on this diagonal 
(the cells (0,1), (3,3), (6,5) and so on)
have the same dimension, in the limit $\epsilon\rightarrow0$, 
the same as the field $\Phi_{1,3}$ which is the base 
of this diagonal.

The second $1/\epsilon^{2}$ diagonal is the one which starts with the field $\Phi_{1,7}$, the cell $l=0$, $k=3$, 
having the dimension 
\beq
\Delta_{1,7}=15\,,
\label{eq3.5}
\eeq
in the limit. All the fields on this diagonal have the same dimension, that of the field $\Phi_{1,7}$ at the base.

And so on. The $1/\epsilon^{2}$ fields which are placed 
on the left of the first diagonal are the exact copies, by reflections with respect to the solide line diagonal in Fig.3, of the $1/\epsilon^{2}$ fields on the right: $(0,1)\rightarrow(2,0)$ and so on. They will not be considered as being different. This is similar to the doubling of primary fields in finite Kac tables, 
in the case of minimal models.

The contributions of all the $1/\epsilon^{2}$ fields along 
the first diagonal in Fig.3, to the expansion over the intermediate channels in the equation (\ref{eq2.36}),  
are all the same.
In particular, the conformal block functions 
$F_{2l+1,2k+1}(z)$ are the same for all the fields along
 the diagonal: they all have equal coefficients $\{k_{i}\}$,
\beq
F_{2l+1,2k+1}(z)=1+k_{1}z+k_{2}z^{2}+k_{3}z^{3}+... \,\,,
\label{eq3.6}
\eeq
in the limit $\epsilon\rightarrow 0$.

This is but with one exception, the operator algebra constants $(D^{(2l+1,2k+1)}_{\sigma,\sigma})^{2}$ are different for the $1/\epsilon^{2}$ fields along the  diagonal. With some use of Mathematica one finds that
\beq
(D^{\Phi^{(1)}_{(s)}}_{\sigma,\sigma})^{2}=(D^{\Phi^{(1)}_{(0)}}_{\sigma,\sigma})^{2}\times\frac{1}{(2s+1)^{2}}
\label{eq3.7}
\eeq
Here $\Phi^{(1)}_{(s)}$ is the $1/\epsilon^{2}$ field, along the first diagonal, number $s$:
\bea
\Phi^{(1)}_{(0)}=\Phi_{1,3}, \mbox{the cell (0,1), the base} \nn\\
\Phi^{(1)}_{(1)}=\Phi_{7,7},\,\, \mbox{the cell}\,\, (3,3)\nn\\
\Phi^{(1)}_{(2)}=\Phi_{13,11},\,\, \mbox{the cell}\,\, (6,5)
\label{eq3.8}
\eea
and so on.

Then, in view of the above information, resuming the contributions of the $1/\epsilon^{2}$ fields along the first diagonal, to the expansion in (\ref{eq2.36}), amounts to resuming the coefficients (\ref{eq3.7}) over $s$:
\bea
\sum^{\infty}_{s=0}(D^{\Phi^{(1)}_{(s)}}_{\sigma,\sigma})^{2}=(D^{\Phi^{(1)}_{(0)}}_{\sigma,\sigma})^{2}\times\sum^{\infty}_{s=0}\frac{1}{(2s+1)^{2}}=(D^{\Phi^{(1)}_{(0)}}_{\sigma,\sigma})^{2}\times\frac{\pi^{2}}{8},\nn\\
(D^{\Phi^{(1)}_{(0)}}_{\sigma,\sigma})^{2}_{eff}=(D^{\Phi^{(1)}_{(0)}}_{\sigma,\sigma})^{2}\times\frac{\pi^{2}}{8}
\label{eq3.9}
\eea
Here $(D^{\Phi^{(1)}_{(0)}}_{\sigma,\sigma})^{2}_{eff}$ is the effective constant for the first diagonal.

For the second diagonal, with the $\Phi^{(2)}_{(0)}=\Phi_{1,7}$ (cell $l=0$, $k=3$) at the base, one finds:
\bea
(D^{\Phi^{(2)}_{(s)}}_{\sigma,\sigma})^{2}=(D^{\Phi^{(2)}_{(0)}}_{\sigma,\sigma})^{2}\times\left(\frac{3(s+1)}{(2s+1)(2n+3)}\right)^{2}\nn\\
(D^{\Phi^{(2)}_{(0)}}_{\sigma,\sigma})^{2}_{eff}=(D^{\Phi^{(2)}_{(0)}}_{\sigma,\sigma})^{2}\times\sum_{s=0}^{\infty}\left(\frac{3(s+1)}{(2s+1)(2n+3)}\right)^{2}=(D^{\Phi^{(2)}_{(0)}}_{\sigma,\sigma})^{2}\times\frac{9\pi^{2}}{64}
\label{eq3.10}
\eea

For the third diagonal, the field $\phi^{(3)}_{(0)}=\Phi_{1.11}$ (cell $l=0$, $k=5$) at the base:
\bea
(D^{\Phi^{(3)}_{(s)}}_{\sigma,\sigma})^{2}=(D^{\Phi^{(3)}_{(0)}}_{\sigma,\sigma})^{2}\times\left(\frac{15}{2}\times\frac{(s+1)(s+2)}{(2n+1)(2n+3)(2n+5)}\right)^{2}\nn\\
(D^{\Phi^{(3)}_{(0)}}_{\sigma,\sigma})^{2}_{eff}=(D^{\Phi^{(3)}_{(0)}}_{\sigma,\sigma})^{2}\times\frac{2475\pi^{2}}{16384}
\label{eq3.11}
\eea

Still one more diagonal, the fourth one, $\Phi^{(4)}_{(0)}=\Phi_{1.15}$ (cell $l=0$, $k=7$) at the base:
\bea
(D^{\Phi^{(4)}_{(s)}}_{\sigma,\sigma})^{2}=(D^{\Phi^{(4)}_{(0)}}_{\sigma,\sigma})^{2}\times\left(\frac{35}{2}\times\frac{(s+1)(s+2)(s+3)}{2s+1)(2s+3)(2s+5)(2s+7)}\right)^{2}\nn\\
(D^{\Phi^{(4)}_{(0)}}_{\sigma,\sigma})^{2}_{eff}=(D^{\Phi^{(4)}_{(0)}}_{\sigma,\sigma})^{2}\times\frac{20825\pi^{2}}{131072}
\label{eq3.12}
\eea

The generalisation is evident. The numerical coefficient 
of the $s$ - dependent factor, like 3, $\frac{15}{2}$, 
$\frac{35}{2}$, in the above formulas, is such that this factor, on the whole, becomes equal to 1 when $s=0$.

One could provide the whole series, for the expansion (\ref{eq2.36}), where the summation which remains is that over the diagonals, with coefficients $(D)^{2}$ effective and the conformal block functions $|F(z)|^{2}$ for the fields at the base, for each diagonal.

But here the numerical observation comes. One finds that the ratio of the first two $(D)^{2}_{eff} $ coefficients:
\beq
\frac{(D^{\Phi^{(2)}_{(s)}}_{\sigma,\sigma})^{2} _{eff}}{(D^{\Phi^{(1)}_{(0)}}_{\sigma,\sigma})^{2}_{eff}}\sim\frac{(D^{\Phi^{(2)}_{(0)}}_{\sigma,\sigma})^{2}}{(D^{\Phi^{(1)}_{(0)}}_{\sigma,\sigma})^{2}}\equiv\frac{(D^{\Phi_{1,7}}_{\sigma,\sigma})^{2}}{(D^{\Phi_{1,3}}_{\sigma,\sigma})^{2}}
\label{eq3.13}
\eeq
-- this ratio is extremely small. With the use of Mathematica, one finds:
\beq
(D^{\Phi_{1,3}}_{\sigma,\sigma})^{2}=-\frac{25}{9437184}\times\frac{1}{\epsilon^{2}}\simeq-\frac{2.6491\times10^{-6}}{\epsilon^{2}}
\label{eq3.14}
\eeq
\beq
(D^{\Phi_{1,7}}_{\sigma,\sigma})^{2}\simeq-\frac{8.0604\times10^{-38}}{\epsilon^{2}}
\label{eq3.15}
\eeq
Their ratio is equal to:
\beq
\frac{(D^{\Phi_{1,7}}_{\sigma,\sigma})^{2}}{(D^{\Phi_{1,3}}_{\sigma,\sigma})^{2}}\simeq3.0427\times10^{-32}
\label{eq3.16}
\eeq
which makes it reasonable to neglect the contribution of other diagonals, by keeping in (\ref{eq2.36}) just the contribution of the first one, with the effective constant
\beq
(D^{\Phi_{1,3}}_{\sigma,\sigma})^{2}_{eff}=(D^{\Phi_{1,3}}_{\sigma,\sigma})^{2}\times\frac{\pi^{2}}{8}
\label{eq3.17}
\eeq
and the corresponding conformal block function of the base field $F_{1,3}(z)$.

We have calculated, with the standard technique based on the Virassoro algebra, the first six coefficients of this function:
\bea
F_{1,3}(z)=1+k_{1}z+k_{2}z^{2}+k_{3}z^{3}+k_{4}z^{4}+k_{5}z^{5}+k_{6}z^{6}+...\nn\\
k_{1}=1\nn\\
k_{2}=\frac{16855}{18432}\simeq0,91444\nn\\
k_{3}=\frac{7639}{9216}\simeq0,82888\nn\\
k_{4}=\frac{769359145}{1019215872}\simeq0,75485\nn\\
k_{5}=\frac{235218217}{339738624}\simeq0,69235\nn\\ 
k_{6}=\frac{72083915765407}{112717121716224}\simeq0,63951
\label{eq3.18}
\eea

Within the above described approximation (neglecting the higher diagonals) one finds the following expression for the four spins function:
\bea
\lim_{\epsilon\rightarrow0}\epsilon^{2}<\sigma(\infty)\sigma(1)\sigma(z,\bar{z})\sigma(0)>\nn\\
\simeq\lim_{\epsilon\rightarrow0}\epsilon^{2}\frac{1}{|z|^{4\Delta_{\sigma}-2\Delta_{1,3}}}(D^{(1,3)}_{\sigma,\sigma})^{2}_{eff}|F_{1,3}(z)|^{2}\nn\\
=-\frac{25}{9437184}\frac{\pi^{2}}{8}|z|^{91/24}
\times|F_{1,3}(z)|^{2}
\label{eq3.19}
\eea

We remind that this function has been defined for the normalized spin operators, $<\sigma(z,\bar{z})\sigma(0)>=1/|z|^{4\Delta_{\sigma}}$.

$(D^{(1,3)}_{\sigma,\sigma})^{2}$ is negative because the norm squared $(N_{1,3})^{2}$ in eq.(\ref{eq2.44}) is negative, the model is not unitary.
We remind that the norm squared of CG vertex operators is defined by the equation (\ref{eq2.49}). One can check
that $(N_{1,3})^{2} = -\frac{225}{64}$.

\section{Discussion.}

As has already been argued in \cite{ref9}, to realize on the lattice
the four spins correlation function, of the conformal theory, might be difficult.
The reason is that the lattice spins are, in general, linear combinations
of the conformal theory spin-like operators.
So that the lattice four spins correlation function will,
in general, be a linear combination of different conformal theory
four points correlation functions, of spin-like operators.

When doing different linear combinations of lattice defined four points connectivities,
the number of possibilities is limited, on the lattice.

One may think of doing the analysis in the opposite direction: taking,
in the conformal theory context, a linear combination of different spin-like operators,
with undefined coefficients, calculate the corresponding four points functions and then try
to fine tune the undefined 
coefficients so as to match a particular lattice defined 
four points connectivity.

To realise this approach might be easier in the context 
of a particular minimal model, where the number 
of spin-like operators, all known, is finite. Still it requires some work to be done,
in the conformal theory context, 
to calculate a number of four points functions which arise.

\vskip1cm

{\bf Acknowledgments.}

I am grateful to Marco Picco and Raoul Santachiara
for numerous useful  discussions.

I am also grateful to my wife Valentina Dotsenko for the very useful help
with calculations by Mathematica and equally for the help with preparation
of the manuscript.


\begin{thebibliography}{99}


\bibitem{ref1a}       G. Delfino and J. Viti, 
                     J. Phys. A: Math. Theor. {\bf 44} (2011) 032001
                     
\bibitem{ref1b}       G. Delfino and J. Viti, 
                      Nucl. Phys.  {\bf B 852} (2011) 149                  

\bibitem{ref1c}       M. Picco, R. Santachiara, J. Viti and G. Delfino, 
                     Nucl. Phys.  {\bf B 875} (2013) 719  



\bibitem{ref1e}     J. L. Jacobsen,  H. Saleur (2018) arXiv:1809.02191

\bibitem{ref1f}      M. Picco, Sylvain Ribault and R. Santachiara (2019)
                      arXiv:1906.02566



\bibitem{ref1}        Vl. S. Dotsenko,
                     Nucl. Phys. {\bf B 235} (1984) 54               

\bibitem{ref2}       Vl. S. Dotsenko and V. A. Fateev,
                     Nucl. Phys.  {\bf B 240} (1984) 312

\bibitem{ref3}        A. A. Belavin, A. M. Polyakov, A. B. Zamolodchikov,
                     Nucl. Phys. {\bf B 241} (1984) 333         

\bibitem{ref4}       Vl. S. Dotsenko and V. A. Fateev,
                     Nucl. Phys.  {\bf B 251} (1985) 691

\bibitem{ref5}       D. Friedan, Z. Qiu and S. Shenker,
                      Phys. Rev. Lett. {\bf 52} (1984) 1575

\bibitem{ref6}       Al. B. Zamolodchikov, 
                     Theor. Math. Phys. {\bf 142} (2005) 183
                      
\bibitem{ref8}       Vl. S. Dotsenko, Nucl. Phys. {\bf B 907}                       
                      (2016) 208
\bibitem{ref9}       Vl. S. Dotsenko, Nucl. Phys. {\bf B 911}                       
                      (2016) 712



                      


                    
                               
\end{thebibliography}
\end{document}